\documentclass[a4paper,10pt]{article}
\usepackage[pdftex]{graphicx,hyperref}
\usepackage{amsmath,amsfonts,amssymb}
\usepackage{wasysym}
\usepackage{fullpage}
\usepackage{enumerate}
\usepackage{sidecap}
\usepackage{lscape}
\usepackage{authblk}
\usepackage{appendix}
\usepackage{setspace}
\title{\bf{Time varying networks and the weakness of \\ strong ties}}
\author[1,2]{M\'arton Karsai}
\author[1]{Nicola Perra}
\author[1,3,4]{Alessandro Vespignani \thanks{a.vespignani@neu.edu}}
\affil[1]{Laboratory for the Modeling of Biological and Socio-technical Systems, Northeastern University, Boston MA 02115 USA}
\affil[2]{Department of Biomedical Engineering and Computational Science, School of Science, Aalto University, P.O. Box 12200, FI-00076}
\affil[3]{Institute for Scientific Interchange Foundation, Turin 10133, Italy}
\affil[4]{Institute for Quantitative Social Sciences at Harvard University, Cambridge, MA, 02138}

\begin{document}

\maketitle

\begin{abstract}
In most social and information systems the activity of agents generates rapidly evolving time-varying networks. The temporal variation in networks' connectivity patterns and the ongoing dynamic processes are usually coupled in ways that still challenge our mathematical or computational modelling. Here we analyse a mobile call dataset and find a simple statistical law that characterize the temporal evolution of users' egocentric networks. We encode this observation in a reinforcement process defining a time-varying network model that exhibits the emergence of strong and weak ties. We study the effect of time-varying and heterogeneous interactions on the classic rumour spreading model in both synthetic, and real-world networks. We observe that strong ties severely inhibit information diffusion by confining the spreading process among agents with recurrent communication patterns. This provides the counterintuitive evidence that strong ties may have a negative role in the spreading of information across networks\footnote{Journal reference: M. Karsai, N. Perra and A. Vespignani, \textit{Scientific Reports} \textbf{4}, 4001 (2014).}.
\end{abstract}

In the last ten years the access to high resolution datasets from mobile devices, communication, and pervasive technologies has propelled a wealth of developments in the analysis of large-scale networks \cite{lazerscience2009,alex12-1,butts08-1,Newman2010Networks}. A specific  effort has been devoted to characterize how network's  structure influences the behaviour of dynamical processes evolving on top of them, an extremely important question for the understanding and modelling of the spreading of ideas, diseases, informations, and many others dynamical phenomena \cite{boccaletti06-1,Christakis2007,Barrat2008Dynamical,aral2009,Volz2009}. However, the large majority of approaches put forth so far uses a time-aggregated representation of network's interactions, neglecting the time-varying nature of real systems connectivity patterns. This approximation is extremely convenient for the sake of mathematical and computational analysis, but it is prone to introduce strong biases in the description of the dynamical  processes occurring on the network~\cite{alex12-1,morris93-1,morris07-1,clauset07,Rocha:2010,Isella:2011,Stehle:2011nx,Karsai2011Small,Miritello2011Dynamical, albert2011sync,Parshani:2010,Bajardi:2011, consensus_temporal_nrets_2012,starnini_rw_temp_nets,Perra2012Activity,Perra2012Walking,ribeiro12-1,liu13-1,lambiotte12-1,pfitzner13-1}. Indeed, the concurrency, and time ordering of interactions, are crucial in a correct description of network's processes~\cite{morris95-1,morris07-1,Toro2007,ButtsScience2012,Rocha2012Epidemics,Perra2012Activity}.

The characterization and modelling of time-varying networks are still open and active areas of research \cite{Handcock2003,Holme2012Temporal}. In this context, relational event-based network analysis enable to model network dependent, time-stamped event data \cite{butts08-1} as well as human, and organizational interactions \cite{brandes09-1,nooy08-1}. Appropriate dyadic level statistics govern the rate at which actors send out communications to their neighbours encoding traditional network structures as well as actor level attributes or even the history of actor level events for the sender. A simplification of this framework has been recently proposed by the activity-driven generative algorithm for time-varying networks~\cite{Perra2012Activity}.  This approach is based on the activity potential, a time invariant function characterizing agents' interactions. This class of models generates activity-driven networks that provides a simplified picture of highly dynamical networks~\cite{Perra2012Activity,Perra2012Walking,ribeiro12-1,liu13-1}. The activity-driven framework has considered only memoryless generative processes so far. At each time step, nodes select their partners with a uniform probability. The model thus neglects the heterogeneous nature of individuals' social interactions. Indeed, in real social systems, agents have strong ties defined as connections that are frequently repeated, and weak ties signalling occasional interactions.  The heterogeneity of social ties is a key ingredient of social networks and plays a crucial role on diffusion processes~\cite{Onnela2007Structure}. However, a full understanding of the mechanism driving their formation and their effects on dynamical phenomena, explicitly considering the network's time-varying nature, is still missing. 

In this paper we propose an extension of the activity-driven framework to model and capture the emergence of heterogeneous ties in social networks.  We perform a thorough analysis of a large-scale mobile phone-call (MPC) dataset containing time-stamped communication events of more than six million individuals (for detailed description see Methods). In this system, the interaction dynamics of a node (ego) can be explained by introducing simple memory effects encoded in a non-Markovian reinforcement process. The introduction of this mechanism in the activity-driven model allows capturing the evolution of the egocentric network of each actor in the system. Within this new framework we study a family of information propagation processes, namely the rumour spreading model~\cite{Daley1964Epidemics,Maki1973Mathematical}. We tackle the case in which the dynamics of contacts and the spreading process are acting on the same time-scale. Interestingly, both in synthetic and real time-varying networks we find that memory hampers the rumour spreading process. Strong ties have an important role in the early cessation of the rumor diffusion by favouring interactions among agents already aware of the gossip. The celebrated Granovetter conjecture that spreading is mostly supported by weak ties \cite{granovetter73-1}, goes along  with a negative effect of strong ties. In other words, while favouring locally the rumor spreading, strong ties have an active role in confining the process for a time sufficient to its cessation. 

\section*{Results}

\begin{figure}[ht]
\begin{center}
\includegraphics[width=0.55\textwidth,angle=0]{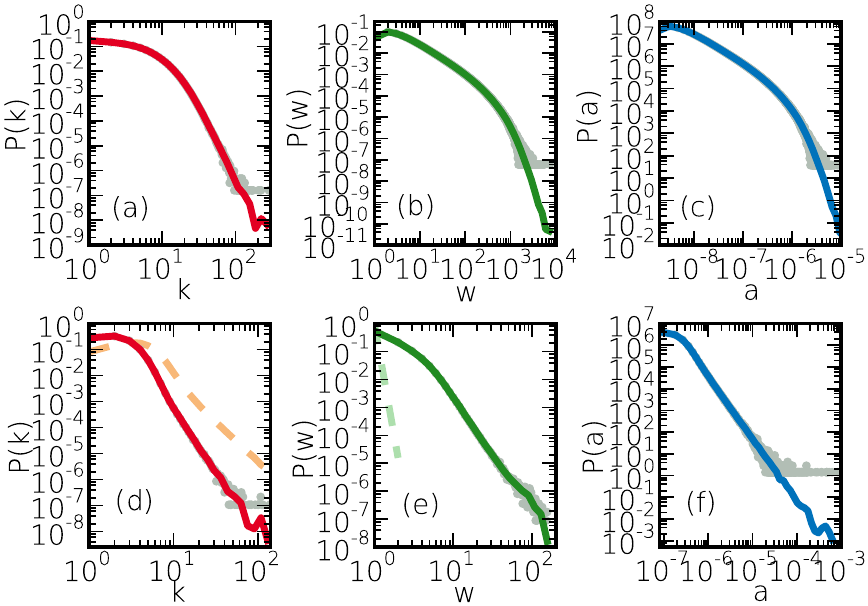}
\caption{Distributions of the characteristic measures of the aggregated MPC network, and activity-driven networks. In panels (a), and (d) we plot the degree distributions. In panels (b), and (e) we plot the weight distributions. Finally, in panels (c), and (f) we plot the activity distributions. In each figure grey symbols are assigning the original distributions while coloured symbols are denoting the same distributions after logarithmic binning. Measured quantities in MPC sequences were recorded for $182$ days (see Methods). In panels (d), (e), and (f) solid lines are assigned to the distributions induced by the reinforced process, while dashed lines denote results of the original memoryless process. Model calculations were performed with parameters $N=10^6$, $\epsilon=10^{-4}$ and $T=10^4$.}
\label{fig1}
\end{center} 
\end{figure}

\begin{figure}[ht]
\begin{center}
\includegraphics[width=0.9\textwidth,angle=0]{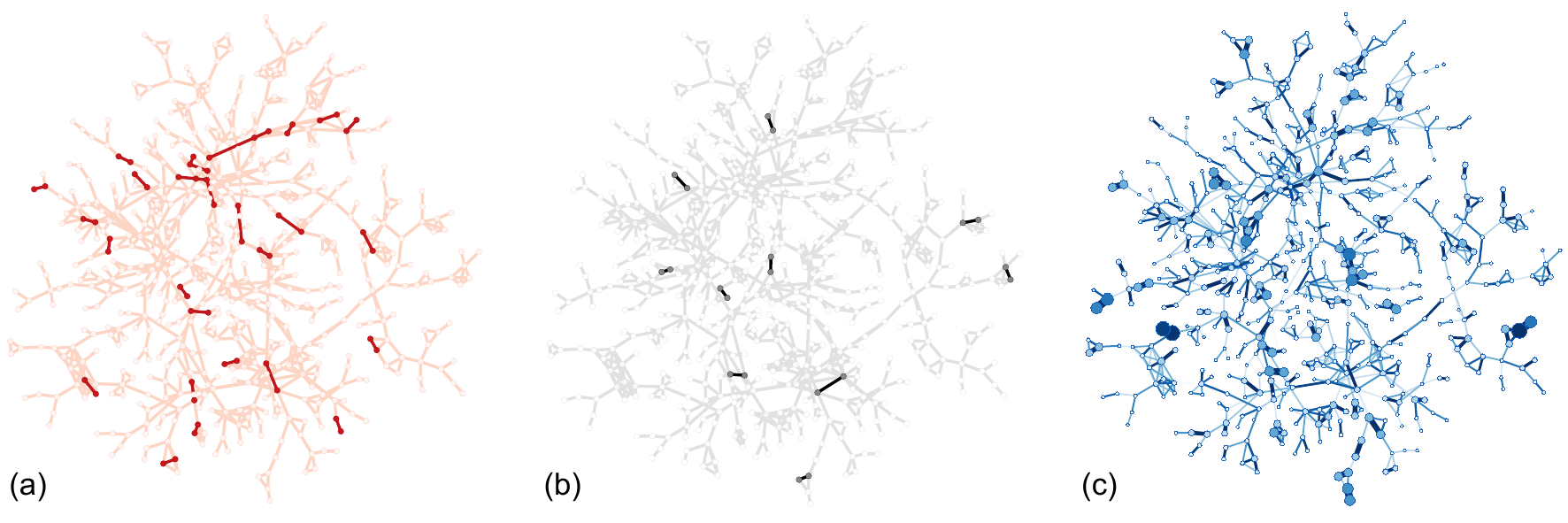}
\caption{ Dynamics of the MPC network. Panels (a), and (b) show calls within $3$ hours between people in the same town in two different time windows. Panel (c) presents the total weighted social network structure, which was recorded by aggregating interactions during $6$ months. Node size and colors describe the activity of users, while link width and color represent weight.}
\label{fig2}
\end{center} 
\end{figure}

We focus on a prototypical large scale communication network where  mobile phone users are nodes and the calls among them links. The common analysis framework for such systems neglects the temporal nature of the connections in favour of time-aggregated representations. In these representations, the degree $k$ of a node indicates the total number of contacted individuals, while the weight of a link $w$ (the strength of the tie) the total number of calls between the pair of connected nodes. The distributions of these quantities are shown in Fig.\ref{fig1}.a, and b. Interestingly, they are characterized by heavy-tailed distributions. Although, the study of the time-aggregated network provides basic information about its structure, it cannot inform us on the processes driving its dynamics. This intuition is clearly exemplified in Fig.\ref{fig2}.a and b. These figures show two snapshots of the network at different times covering few hours of calls in a town. The two plots capture dynamical interaction patterns not visible from the aggregated network representation (Fig.\ref{fig2}.c). 

Here we aim to study and identify the mechanisms driving the evolution, and dynamics of the egocentric networks (egonets) of the global network.
Egonets were thoroughly investigated earlier in psychology and sociology \cite{Wasserman1994Social,Barrett2002Human, Degenne1999Introducing}. Some other characteristics have been recently mapped out with the availability of large-scale data \cite{Fisher2005Using,Newman2003Egocentered,Karsai2012Correlated,goncalves:twitter,miritello12-1}. We tackle this problem from a different angle focusing on the activity rate, $a$, that allows describing the network evolution beyond simple static measures. It is defined as the probability of any given node to be involved in an interaction at each unit time. The activity distribution is also heavy-tailed (see Fig.\ref{fig1}.c), but contrary to degree and weight, is a time invariant property of individuals~\cite{Perra2012Activity}. It does not change by using different time aggregation scales~\cite{Perra2012Activity,ribeiro12-1}. This quantity is the basic ingredient of the activity-driven modelling framework~\cite{Perra2012Activity}. Here we extend this approach by identifying, and modelling another crucial component: the  memory of each agent. We encode this ingredient in a simple non-Markovian reinforcing mechanism that allows to reproduce with great accuracy the empirical data.   

\subsection{Egocentric network dynamics.}

\begin{figure}[ht]
\begin{center}
\includegraphics[width=0.55\textwidth,angle=0]{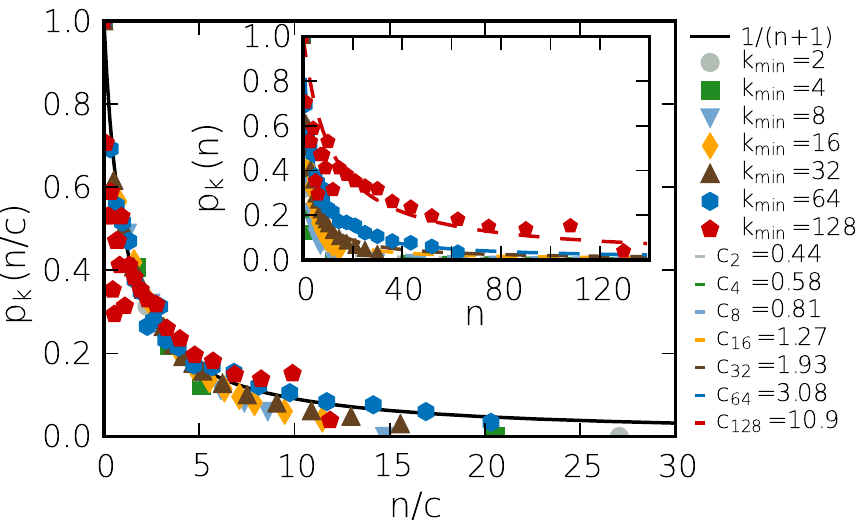}
\caption{The $p_k(n)$ probability functions calculated for different degree groups in the MPC network. In the inset, symbols show the averaged $p_k(n)$ for groups of nodes with degrees between the corresponding $k_{min}...k_{min}^2-1$ values. Continuous lines are the fitted functions of Eq.\ref{eq:pn} with $c$ parameter values showed in the legend. The main panel depicts the same functions after rescaling them using Eq.\ref{eq:pnsc}. The continuous line describes the analytical curve of Eq.\ref{eq:pnsc}.}
\label{fig3}
\end{center} 
\end{figure}

In general, social networks are characterized by two types of links. The first class describes strong ties that identify time repeated and frequent interactions among specific couples of agents. The second class characterizes weak ties among agents that are activated only occasionally. It is natural to assume that strong ties are the first to appear in the system, while weak ties are incrementally added to the egonet of each agent~\footnote{Note that here we are not considering additional dynamic processes such as changes in social status, ageing, or permanent breaking of social ties that generally acts on different time scales.}. This intuition has been recently confirmed~\cite{Gautier2012} in a large-scale dataset and indicates a particular egocentric network evolution. In order to quantify it, we measure  the probability, $p(n)$, that the next communication event of an agent having  $n$ social ties will occur via the establishment of a new $(n+1)^{th}$ link. We calculate these probabilities in the MPC dataset averaging them for users with the same degree $k$ at the end of the observation time. We therefore  measure the quantity $p_k(n)$ for the egonets with the same degree $k$ and  $n\leq k$.  The empirical $p_k(n)$ functions for different degree groups are shown in Fig.\ref{fig3} inset (coloured symbols). Interestingly, the probabilities are decreasing with $n$ for each degree class denoting a slow down in the egocentric network evolution. The larger the egocentric network, the smaller the probability that the next communication will be with someone who was not contacted before. Agents have memory. They remember their social ties and tend to repeat interactions on these already established connections.

The empirical growth of the egonet  can be captured by a simple mechanism. We find that the probability that a node, characterized by a social circle of size $n$, will establish a new tie is well fitted by the expression :
\begin{equation}
 p(n)=1-\frac{n}{n+c}=\frac{c}{n+c}.
\label{eq:pn}
\end{equation}
Analogously, the probability of having an interaction with someone who is already in the egocentric network is $n/(n+c)$. Here $c$ is an offset constant depending on the degree class considered. By fitting the function in Eq.\ref{eq:pn} on the empirical data (solid lines in Fig.\ref{fig3} inset) we can determine the corresponding constant $c$ for each degree group (see Table 1 in the Supplementary Materials (SM) for the obtained values). Using the measured $c$  values we can rescale the empirical $p_k(n)$ functions as
\begin{equation}
 p_k(n/c)=1/(n/c+1),
\label{eq:pnsc}
\end{equation}
and collapse the data points of different degree groups on a single curve  (see Fig.\ref{fig3} main panel). This remarkable result suggests that the same mechanism is driving the evolution of the egonets of all individuals independently of their final number of connections.

\subsection{Activity-driven network model with memory.}
The basic activity-driven network model~\cite{Perra2012Activity} considers $N$ nodes, each one assigned with an activity probability per unit time $a_i=\eta x_i$. Here $x_i$ denotes the activity potential drawn from a desired $F(x_i)$ distribution ($x_i\in [ \epsilon ,1]$, $\epsilon$ fixes the minimal value of activity in the system) and $\eta$ is a rescaling factor that fixes the average number of active nodes per unit time to  $\eta \langle x \rangle N$. The generative network process is defined according to the following rules: i) At each discrete time step $t$ the network $G_t$ starts with $N$ disconnected vertices; ii) With probability $a_i \Delta t $ each vertex $i$ becomes active and generates $m$ links that are connected to $m$ other randomly selected vertices; iii) At the next time step $t + \Delta t$, all the edges in the network $G_t$ are deleted. In this formulation inactive nodes can receive connections. Different rules can be easily implemented to model different scenarios~\cite{hoppe13-1}. Without loss of generality we fix the parameters $\eta=1$, $\epsilon=10^{-3}$, and $\Delta t =1$. Furthermore, in order to suit the MPC dataset we set $m=1$, i.e. each call take place between two people. We consider heavy-tailed distributions of activity i.e. $F(x)\propto x^{-\nu}$, that reproduce the behaviour observed in real data for a number of real-world networks~\cite{Perra2012Activity,ribeiro12-1,Cattuto2010Dynamics,Chmiel2009Scaling}. Inspired by measurements in the MCP dataset we set the exponent to $\nu=2.8$  (see Fig.\ref{fig1}.c and f).\\

\begin{figure}[ht]
\begin{center}
\includegraphics[width=0.9\textwidth,angle=0]{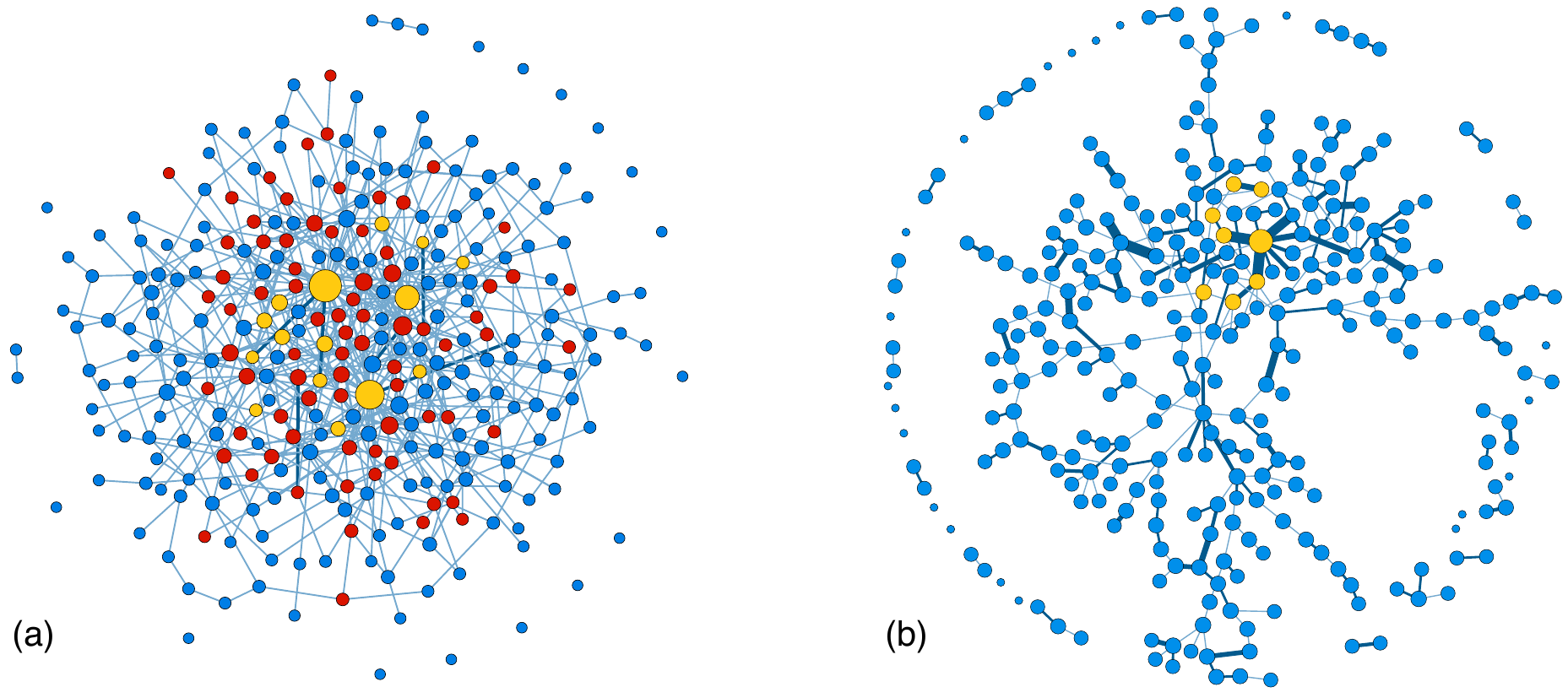}
\caption{Rumour spreading processes in (a) ML and (b) RP activity-driven networks. Node colors describe their states as ignorant (blue), spreader (red) and stifler (yellow). Node sizes, color, and width of edges represent the corresponding degrees and weights. The parameters of the simulations are the same for the two processes: $N=300$, $T=900$, $\lambda=1.0$, and $\alpha=0.6$. The processes were initiated from a single seed with maximum strength.}
\label{fig4}
\end{center} 
\end{figure}

In the basic activity-driven model the network dynamics is memoryless (ML). At each time step all connections previously established  are removed, and the new one are created with no memory of the past. Here  we extend the modelling framework introducing a simple reinforcement process in which nodes keep remembering who they have connected~\cite{Stehle2010Dynamical,Karsai2012Universal,Karsai2012Correlated}. Inspired by the observations in the MCP dataset, we impose a reinforcement mechanism in which an active node with $n$ previously established social ties will contact randomly a new node with probability $p(n)=c/(n+c)$. Otherwise, with probability $1-p(n)=n/(n+c)$ it will interact with a node already contacted, thus reinforcing earlier established social ties. In this case, the selection is done randomly among the $n$ neighbours. This model, that in the following we will denote as RP (reinforcement process), is non-Markovian. Memory is explicitly introduced in the egonetwork dynamics as each node keeps remembering the list of already established ties. We fix $c=1$ for all the nodes and we leave the generalization of the model where this value is correlated with node properties for future studies (indeed we show in the SM how the emerging network properties are changing for different values of $c$).

A side by side comparison of the time-aggregated representations of networks generated by the ML and RP models (using the same parameters) is shown in Fig.\ref{fig4}-a and b.  The ML dynamics (Fig.\ref{fig4}.a) induces an aggregated network with a degree distribution $P(k)\propto k^{-\gamma}$ where $\gamma=\nu$ and a weight distribution decaying exponentially~\cite{Perra2012Activity,starnini13-2}. This is also confirmed by large scale simulation results reported in Fig.\ref{fig1}.d and e (dashed lines). In case of the RP dynamics (Fig.\ref{fig4}.b), the  memory process induces a considerably different structure. These effects are quantified in Fig.\ref{fig1}.d, e, and f (solid lines). We observe a degree distribution that is heavy-tailed but more skewed in the RP model than the ML. This distribution is qualitatively matching  the corresponding empirical measure in Fig.\ref{fig1}.a. Furthermore, the RP model generates heterogeneous weight distributions (see Fig.\ref{fig1}.e solid line) capturing extremely well real data. This is not the case in the ML model where the absence of memory induces exponential weight distributions far from reality (see Fig.\ref{fig1}.e dashed line). The RP dynamics not only induces realistic heterogeneities in the network structure, but also controls the evolution of the macroscopic network components. Indeed, due to the reinforcement mechanism,  the largest connected component (LCC) in RP networks grows considerably slower than in the case of ML models (for illustration see Fig.\ref{fig5}.a). This is an important feature because  dynamical process evolving on time-varying networks will progress with a time-scale that cannot be smaller than the LCC growth time-scale. As consequence, dynamical phenomena taking place on time-varying networks with memory will evolve at a slower rate than in memoryless time-varying networks. In the case of epidemic spreading for example, the memory in individuals' connections patterns shifts the epidemic threshold to larger values, and more in general reduces the final number of infected nodes (see SM for details).

\subsection{Rumour spreading processes on activity-driven networks.}

In order to study the effects of the emergence of strong ties on dynamical processes taking place in the network, we consider the classic rumor spreading process~\cite{Daley1964Epidemics}. In this scheme, each node can be in three possible states; ignorant ($I$), spreader ($S$) or stifler ($R$). We denote the densities of individuals in each state at time $t$ as $i(t)=I(t)/N$, $s(t)=S(t)/N$, and $r(t)=R(t)/N$ accordingly. At $T=0$ everyone is ignorant except the selected single or multiple seeds who are set to be spreaders. At the time of an interaction the states of connected nodes can change by the following rules: (a) $I+S\overset{\lambda}{\rightarrow}2S$ or (b) $S+R\overset{\alpha}{\rightarrow}2R$ or (c) $S+S\overset{\alpha}{\rightarrow}2R$. Here $\lambda$ and $\alpha$ are the transition rates into the states of spreader or stifler accordingly. In all measurement (if it is not noted otherwise) we set $\lambda=1$ and use $\alpha$ as a parameter. We assume that only their ratio matters for the spreading behaviour (supporting results are summarized in SM). Using these rules the spreaders communicate with probability $\lambda$ the rumor to connected agents that become spreaders on their turn. If the spreaders however find that a contacted agent is already aware of the rumours, with probability $\alpha$ loose interest in the rumours and stop spreading it thus becoming a stifler. In the long run the system always reaches an equilibrium state where all spreaders have turned into stiflers, $\partial_t i(t)=0$, and $\partial_t r(t)=0$. Different parameters provide different penetration of the rumor in the network. Interesting quantities to study are the velocity of spreading of the rumor and the total number of agents aware of the gossip at the end of the process (stiflers).

\begin{figure}[ht]
\begin{center}
\includegraphics[width=0.55\textwidth,angle=0]{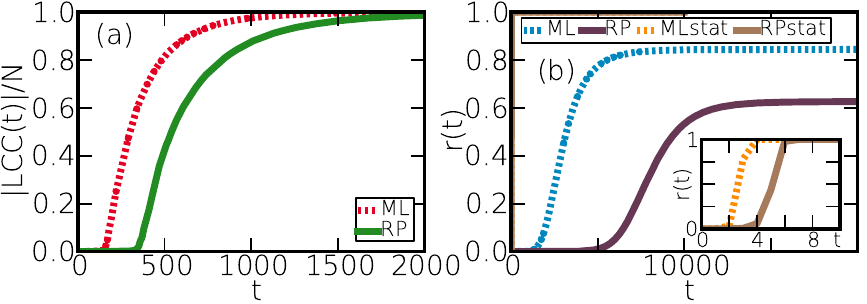}
\caption{In panel (a) we show the sizes of the largest connected components (LCC) as a function of time for time aggregated ML and RP networks. Simulations were run with the same parameters considering $N=10^5$ nodes. In panel (b) we show the stifler $r(t)$ density in rumor spreading simulations in ML (main panel, blue dashed line) and RP (main panel, purple solid line) networks with $N=10^5$ nodes. We set $\lambda=1.0$ and $\alpha=0.6$ and run the simulations for $T=10^5$ time steps. The rumor spreading processes were simulated with the same parameters on aggregated ML (inset, yellow dashed line) and RP (inset, brown solid line) networks integrated for $T$ time steps.}
\label{fig5}
\end{center} 
\end{figure}

Here we are interested in studying the differences on the final contagion densities in networks with or without memory (Fig.\ref{fig5}.b main panel), all other parameters of the rumor spreading model being equal. We set $\lambda=1.0$ and $\alpha=0.6$ and in the case of ML networks at the end of the rumor spreading $\sim 85\%$ of the network is aware of the rumor. Instead, in the RP case the final contagion proportion is only slightly more than $60\%$ of the total nodes. This hampering of the contagion process  is also shown in Fig.\ref{fig4}.a and b for the same set of parameters. The differences are evident not only in the diffusion patterns, but also in the level of contagion. In RP networks, the rumor has spread only locally and reached $6$ nodes other than the seed, while during the same time in the ML networks the information reached $92$ nodes out of $300$.

\begin{figure}[ht]
\begin{center}
\includegraphics[width=0.55\textwidth,angle=0]{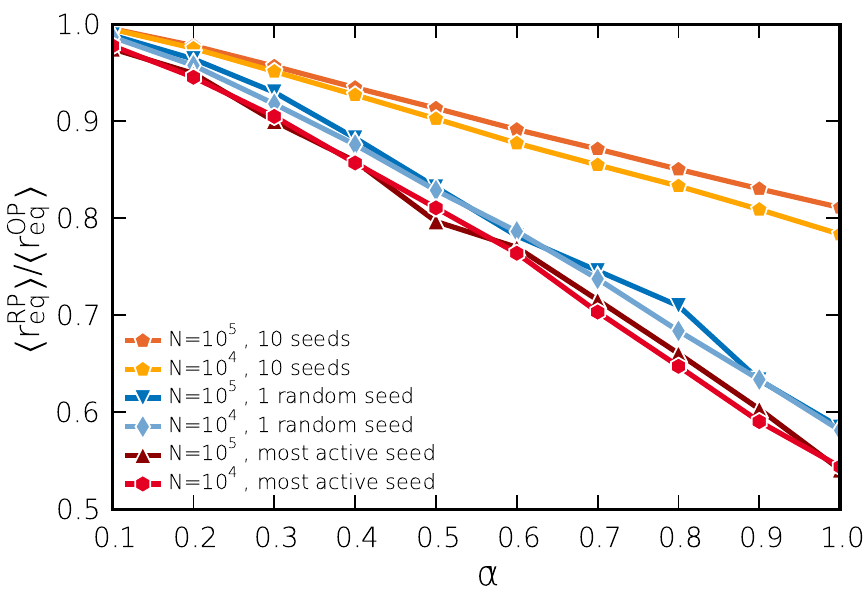}
\caption{ The $r_{eq}^{RP}/r_{eq}^{ML}$ ratios of average stifler densities at equilibrium. The simulations for sizes $10^5$ and $10^4$ were run with various initial conditions (see legend). The averages were calculated at $T=5\times 10^4$ considering only realizations that reached equilibrium after $10^3$ time steps.}
\label{fig6}
\end{center} 
\end{figure}

To investigate in more details rumor spreading processes on different activity-driven models, we perform further simulations using different initial conditions and varying the rumours model parameters. In particular, we initiate the spreading from (i) the most active seed, (ii) one randomly selected seed or (iii) ten random seeds.
We then simulate each process for $T=5\times 10^4$ time steps, and measure the average final proportion of nodes aware of the rumor $\langle r_{eq} \rangle$. 
In each case, we perform $10^3$ (or $10^4$ for smaller systems) simulations in identically parametrized ML and RP networks, where the process  lasts at least $10^3$ steps. To highlight differences arising between the rumor propagation processes evolving on the two network dynamics, we kept $\lambda=1$ and calculate the $\langle r_{eq}^{RP}(\alpha)\rangle / \langle r_{eq}^{ML}(\alpha) \rangle$ ratios as function of $\alpha$. Results in Fig.\ref{fig6} indicate marginal size effects but strong dependence on the initial conditions. All corresponding ratios are decreasing with $\alpha$, highlighting increasing differences between the fraction of population reached by the rumor in the two network dynamics. The largest differences are observed for a single initial seed, especially in the case of the most active nodes. These numerical findings can be understood by considering that the rumor spreading and the reinforcement process are occurring on comparable time scales. The reinforcement mechanism induces recurrent interactions that enhance the cessation of rumor spreading by ``pair annihilation" of nodes connected by strong ties. This effect is controlled by $\alpha$ and can induce up to  $\sim 45\%$ relative difference in the population reached by the rumor in the case of the RP model.

In order to understand the biases induced in the dynamical properties of rumor spreading processes by the time aggregated representation of the networks, we consider topologies generated by a time-aggregated view of ML and RP models (see Fig.\ref{fig5}.b inset) and compare the results with their time-varying counterparts (see Fig.\ref{fig5}.b main panel).  The results obtained show striking differences between the velocity of spreading. Indeed, the time for the rumor to reach a consistent fraction of nodes varies four orders of magnitudes in the two cases, with a very slow spreading dynamics in time-varying networks. Interestingly, this behaviour is general to all spreading processes. The observed results indicate a clear difference between the dynamical properties of processes taking place on time aggregated or time resolved networks. Our findings confirm that, when the time-scale of the processes is comparable with the evolution of the network, static representations of the system might introduce strong biases on the correct characterization of the phenomenon.

\subsection{Rumour spreading processes on real time-varying networks.}

\begin{figure}[ht]
\begin{center}
\includegraphics[width=0.55\textwidth,angle=0]{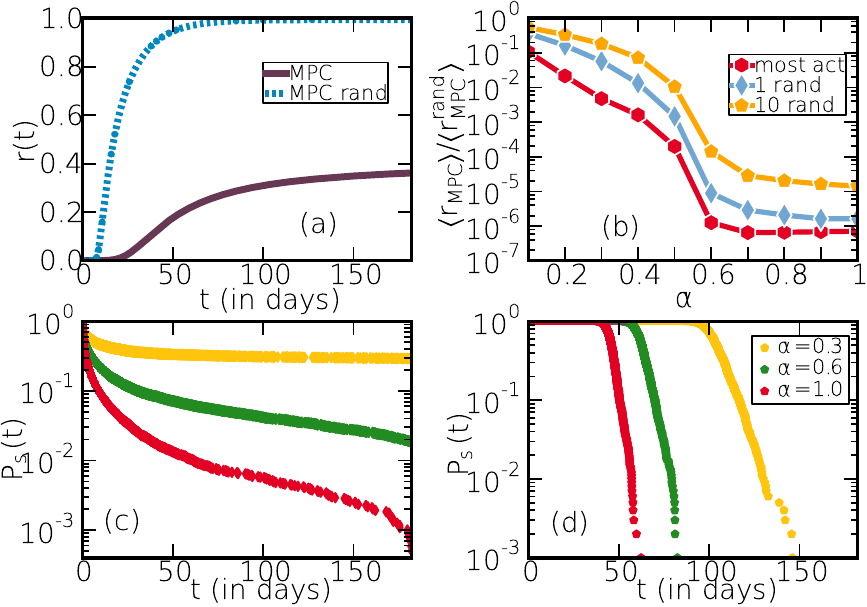}
\caption{In panel (a) we show the stifler $r(t)$ density in data-driven rumor spreading simulations run on top of the MPC dataset (purple solid line) and the MPC null model (blue dashed line) with $\alpha=0.1$. Panel (b) depicts the $r_{MCN}/r_{MCN}^{rand}$ ratios of average stifler densities at equilibrium. Simulations of panels (a) and (b) were run with various initial conditions (see legend) and averaged over $10^3$ realizations. In panels (c) and (d) we plot the surviving probability, $P_s(t)$, of rumor spreading processes initiated from a single random seed in the real MPC sequence and the MPC null model respectively. Probability values of panels (c) and (d) were averaged over $10^4$ realizations.}
\label{fig7}
\end{center} 
\end{figure}

To verify the picture emerging from synthetic time-varying networks, we study the properties of rumor spreading processes in a  real world time-varying system. In particular, we consider the MPC dataset and simulate the rumor spreading by using the actual sequence of calls (for more details see Methods) \cite{Karsai2011Small}. At the same time to directly contrast the role of memory and repeated interactions we defined a random null model defined by keeping the caller of each event as it appears in the MPC dataset, but selecting a callee randomly. In this way, we obtain a sequence recovering the original activities and shuffled egocentric networks. Furthermore, inter-event correlations are removed. The corresponding simulation results in Fig.\ref{fig7}.a shows a clear difference in the speed of spreading and final density of stifler nodes. While in the null model everyone becomes stifler at the end of the simulation, by using the original interaction sequences less than $40\%$ of the network is aware of the rumor. This effect is even more clear in Fig.\ref{fig7}.b where their relative difference is rapidly increasing and becomes several orders of magnitude larger for larger $\alpha$ values. Different initial conditions are playing similar roles as we observed in synthetic networks. The effect of memory and repetitive interactions are the strongest if we initiate the rumor from the most active individual. We observed similar but weaker effects selecting a single or multiple random seeds. 

We also measure the surviving probability $P_s(t)$ defined as the probability that a rumor spreading process survives (still contains nodes actively spreading the rumor) up to  time $t$ \cite{Marro1999Nonequilibrium,Satorras2001Epidemic}. We show $P_s(t)$ for different $\alpha$ in Fig.\ref{fig7}.c. The initial scaling of $P_s(t)$ shows that generally the rumor can spread only locally due to repeated interactions occurring on strong links between the seed and its neighbourhood. A very different behaviour emerges if we remove the effect of memory and repeated interactions considering the same quantities measured on the null model (see Fig.\ref{fig7}.d). Here, as the initial effect of repeated interactions vanishes and all realizations survive until the rumor covers the whole network. Note that similar results were obtained for activity-driven model processes presented in the SM. This highlights the significant role of recurrent interactions via strong ties. They play as bottleneck for the information propagation  controlling the global outbreak of rumor spreading phenomena. 

\section*{Discussion}

We have presented the study of a large scale dataset of social interactions via mobile phone calls. We provided a simple empirical characterization of the effects of memory in its microscopic dynamical evolution.  Considering the empirical evidences, we defined a novel generative model for time-varying networks with memory. The model mirrors many of the structural properties observed in
the real network, like degree and weight heterogeneities, and shows the spontaneous emergence of non-trivial connectivity patterns characterized by strong and weak ties.  We characterize the effects of non-Markovian and heterogeneous connectivity patterns on rumor spreading processes. Interestingly, we find that strong ties are responsible for constraining the rumor diffusion within localized groups of individuals.  This evidence points out that strong ties may have an active role in weakening the spreading of information by constraining the dynamical process in clumps of strongly connected social groups. The presented results underline the subtleties inherent to the analysis of dynamical processes in time-varying networks. No one-fits-all picture exists, and a classification of dynamical process behaviour calls for a thorough analysis of each particular processes and networks considered. Furthermore, several extensions of the utilized framework of activity-driven networks are possible. Examples are node-node correlations, heterogeneous dynamics, and bursty behaviour of nodes. The present study thus offers potential avenues for the study of dynamical processes in time-varying networks in complex settings where the memory of agents plays a determinant role in the evolution of the connectivity patterns of the system.

\section*{Methods}
{\bf Dataset.}
The utilized dataset consists of $633,986,311$ time stamped mobile-phone call (MPC) events recorded during $182$ days with $1$ second resolution between $6,243,322$ individuals connected via $16,783,865$ edges. The dataset was recorded by a single operator with $20\%$ market share in an undisclosed European country (ethic statement was issued by the Northeastern University Institutional Review Board). To consider only true social interactions, and avoid commercial communications we used interactions between users who had at least one pair of mutual interactions.

{\bf Data-driven model.}
In data-driven simulations we initiated the rumor spreading from a randomly selected call event of a randomly selected user in the MPC network. We then run the process for the length of the recorded period. When a realization arrived to the last event of the sequence, we used a periodic temporal boundary condition as we continued the process with the first event of the sequence \cite{Karsai2011Small}. However, as the simulations were executed no longer than the recorded time period, no event was used twice during one simulation run.

{\bf Acknowledgments} This work has been partially funded by the NSF CCF-1101743 and NSF CMMI-1125095 awards. We acknowledge support from the FET project MULTIPLEX 317532 and MK acknowledges support from EU’s 7th Framework Program’s FET-Open ICTeCollective project (No. 238597). We thank A.-L. Barab\'asi for the dataset used in this research. We thank B. Ribeiro, R. Burioni and A. Vezzani for useful discussions. 

{\bf Authors contributions} MK, NP, and AV designed the research and participated in the writing of the manuscript. MK analysed the empirical data, performed the numerical calculations and completed the corresponding analytical derivations.

\pagebreak

\hspace{-.2in}\textbf{\large{Supplementary Materials}} \\ \\
\hspace{-.23in}\textbf{\LARGE{Time varying networks and the weakness of strong ties}} \\ \\
 \large{M. Karsai, N. Perra and A. Vespignani}
\section{Measures of egocentric network evolutions by directed communications}
\label{sec:dir}
\begin{figure}[ht!] \centering
  \includegraphics[width=140mm]{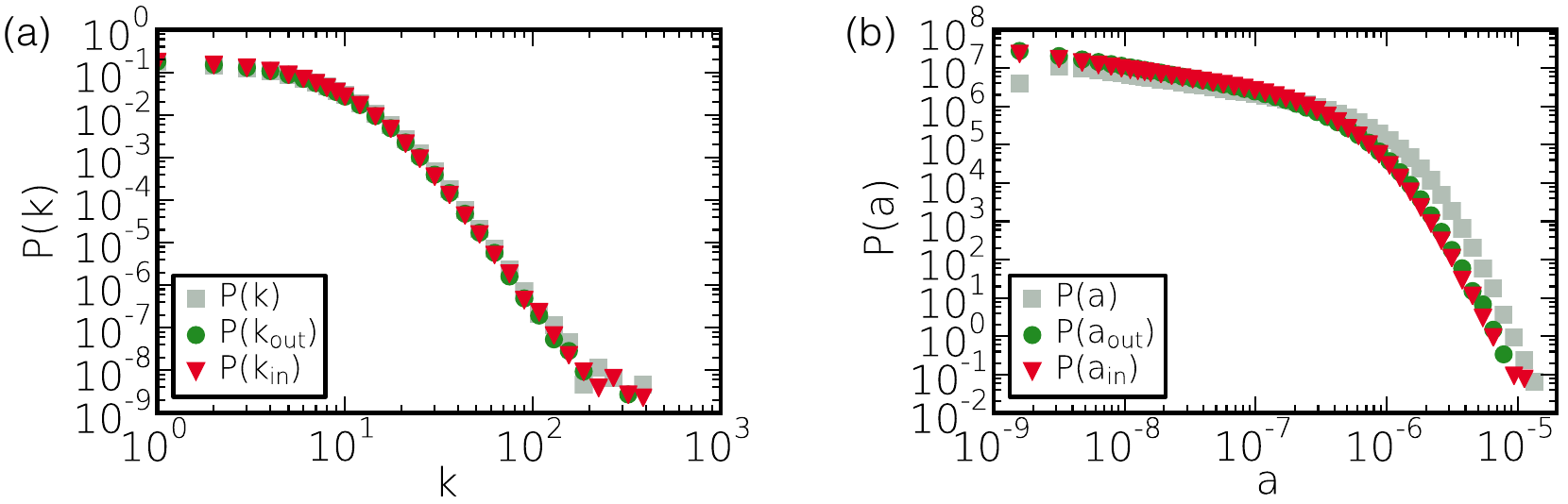}
\caption{Characteristic distributions of networks aggregating directed communications. (a) In- (red) and out (green) degree distributions are compared to the overall undirected $P(k)$ (grey). (b) The incoming and outgoing activity distributions are shown together with the undirected $P(a)$ (with colors similar to figure a).}
\label{fig:SI1}
\end{figure}

\normalsize
In the main text we disclosed strong memory effects in the interactions dynamics of an ego and his/her social circle. To complete our analysis here we repeat all the measurements separately for directed outgoing and incoming call sequences. The characteristic functions as $P(k)$ and $P(a)$ of directed communication are very similar to the undirected case as it is evidenced in Fig.\ref{fig:SI1}.a and b. These counts remain broadly distributed and scaling similarly in all three cases.

To investigate the presence of memory, we categorize every events of an individual into two groups. One group contains actions, which evolve on a link where other events have taken place earlier. Events in this group does not increase the degree of the ego but contribute to the edge weight and activity potential. Events belonging to the other group evolve between the ego and someone else who he/she has never connected before. These events induce links with unit weight, which increase the ego's degree and also incremental to his/her activity. By using this categorization we can measure a conditional probability $p(n)$ that the next event of an individual will be towards one of his $n$ already existing neighbours, or with a new person who he/she has never called before. We measure this probability for different degree groups. We select people with $k$ number of neighbours where $k_{min}\leq k < k_{max}$ and calculate the $p(n)$ function for $n \leq k_{min}$ to assure that each calculated probability value is extracted from the same number of users belonging to the actual group. 

\begin{figure}[ht!] \centering
  \includegraphics[width=145mm]{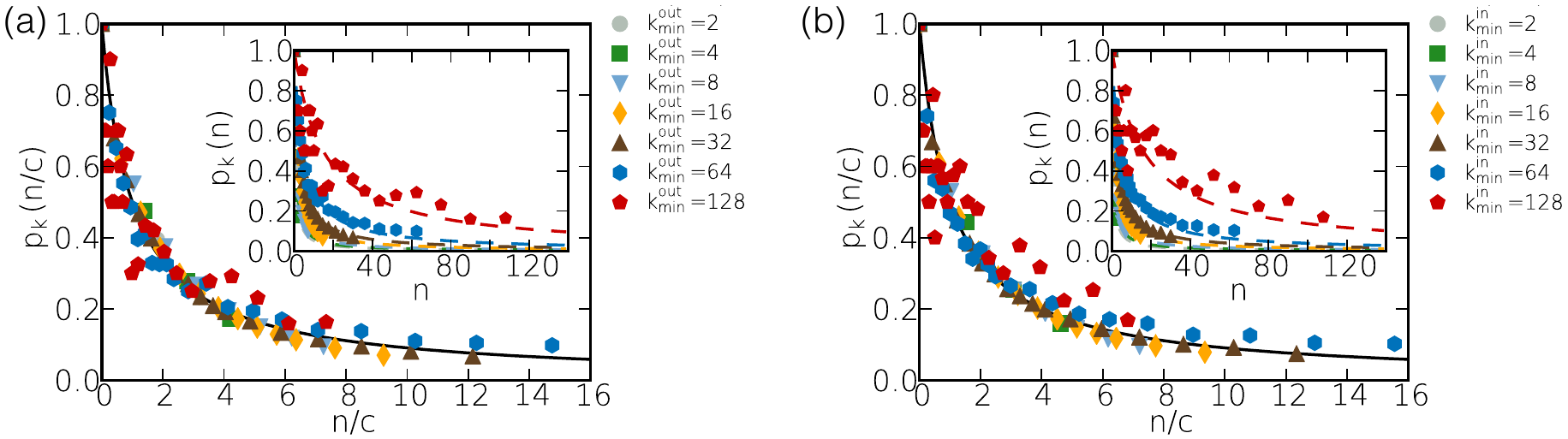}
\caption{Inset: $p_k(n)$ functions for (a) outgoing and (b) incoming call sequences calculated for nodes in different degree groups. We also show fits of $p_k(n)$ functions (defined in Eq.1 in the main text) using different $c$ constant values (see Table.\ref{table:SI1}). Main panel: scaling of $p_k(n)$ and fitting with the universal function defined in Eq.2 (see main text).}
\label{fig:SI2}
\end{figure}

We perform this measurement for directed and undirected communication. Results of undirected communications are reported in the main text while for directed communication sequences are shown in Fig.\ref{fig:SI2}.a and b. They all present very similar behaviour and can be fitted with functions in the form of Eq.1 (see in main text). Fitting results for outgoing and incoming calls are shown in Fig.\ref{fig:SI2}.a and b inset, while their re-scaling (as Eq.2 in main text) is depicted in Fig.\ref{fig:SI2}.a and b main panels accordingly. The corresponding fitted $c$ constants of all directed and undirected sequences are summarized in Table.\ref{table:SI1}.

\begin{table}[h]
\begin{center}
\begin{tabular}{|l|l|l|l|}\hline
$k_{min}..k_{max}$ & $c \pm s_e$ & $c_{out} \pm s_e$ & $c_{in} \pm s_e$ \\ \hline \hline
2..3 & $0.444\pm0.053$ & $0.55317 \pm 0.05973$ & $0.475344 \pm 0.0474$ \\\hline
4..7 & $0.584\pm0.065$ & $0.713296 \pm 0.06143$ & $0.648814 \pm 0.04929$ \\\hline
8..16 & $0.814\pm0.070$ & $0.964174 \pm 0.06003$ & $0.970989 \pm 0.04415$ \\\hline
16..32 & $1.264\pm0.056$ & $1.57125 \pm 0.05505$ & $1.55149 \pm 0.03262$ \\\hline
32..64 & $1.928\pm 0.059$ & $2.46908 \pm 0.05031$ & $2.42959 \pm 0.03998$ \\\hline
64..128 & $3.077\pm0.109$ & $4.23814 \pm 0.2622$ & $4.01782 \pm 0.2525$ \\\hline
128..256 & $10.913\pm0.826$ & $14.6949 \pm 1.735$ & $15.8362 \pm 2.278$ \\\hline
\end{tabular}
\end{center}
\caption{The fitted $c$ constants and $s_e$ standard error values for the observed and analyitical $p(n)$ functions for different degree groups in undirected, outgoing and incoming communication sequences.}
\label{table:SI1}
\end{table}

These results evidence that similar memory effects can be detected in directed and undirected communication sequences. They all indicate that the larger one's observed personal social network the larger the probability that he/she will make (receive) a call towards (from) someone who is already in his/her egocentric network.

\section{Degree evolution of reinforced activity driven networks}

\begin{figure}[ht!] \centering
  \includegraphics[width=140mm]{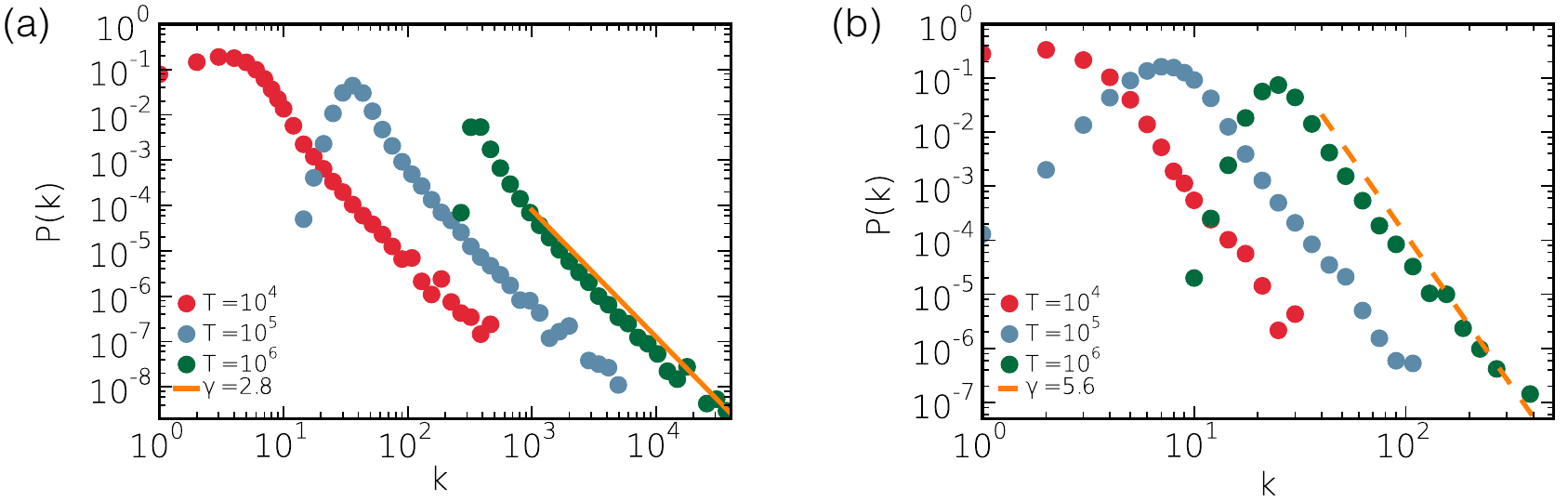}
\caption{Degree distributions of (a) ML and (b) RP activity networks integrated through $T=10^4$, $10^5$ and $10^6$ time steps. The fitting power-law exponent values are (a) $\gamma=2.8$ and (b) $\gamma=5.6$. Common parameters of the simulations were $N=100,000$, $m=1$, $\epsilon=0.0001$.}
\label{fig:DD}
\end{figure}

As it has been shown earlier \cite{Perra2012} the degree distribution of the integrated structure of a memoryless (ML) activity-driven network follow the same functional form as the activity distribution. If the activity distribution scales as $P(a)\sim a^{-\nu}$ then the degree distribution should be $P(k)\sim k^{-\gamma}$ with $\nu=\gamma$. This was shown analytically in \cite{Perra2012} and is confirmed by simulation results in Fig.\ref{fig:DD}.a where $P(k)$ evolves with the same exponent value $\gamma=\nu=2.8$ as $P(a)$ was characterized. The distributions there also show strong finite-size effects. If the simulated network size is finite and we increase the integration time the networks become more and more connected approaching to a fully connected graph. This effect causes that small degree nodes are not presented for large integration time.

Networks generated by reinforced activity-driven processes (RP) also evolve with heterogeneous degrees and with similar finite size effects as above. However, in this case the relation between $\gamma$ and $\nu$ is somewhat different. Here the egocentric network evolution is controlled by the reinforced interactions. Egonets evolve slower as interactions of agents are reinforced to take place on already established links. This effect induce reduced degree heterogeneities with exponents $\gamma$ larger than the one characterizing the activity distribution. It is visible in Fig.\ref{fig:DD}.a and b where $\gamma$ of the evolving ML and RP networks are strikingly different even the activity exponent $\nu=2.8$ and any other parameters were chosen to be the same for both processes.

\begin{figure}[ht!] \centering
  \includegraphics[width=140mm]{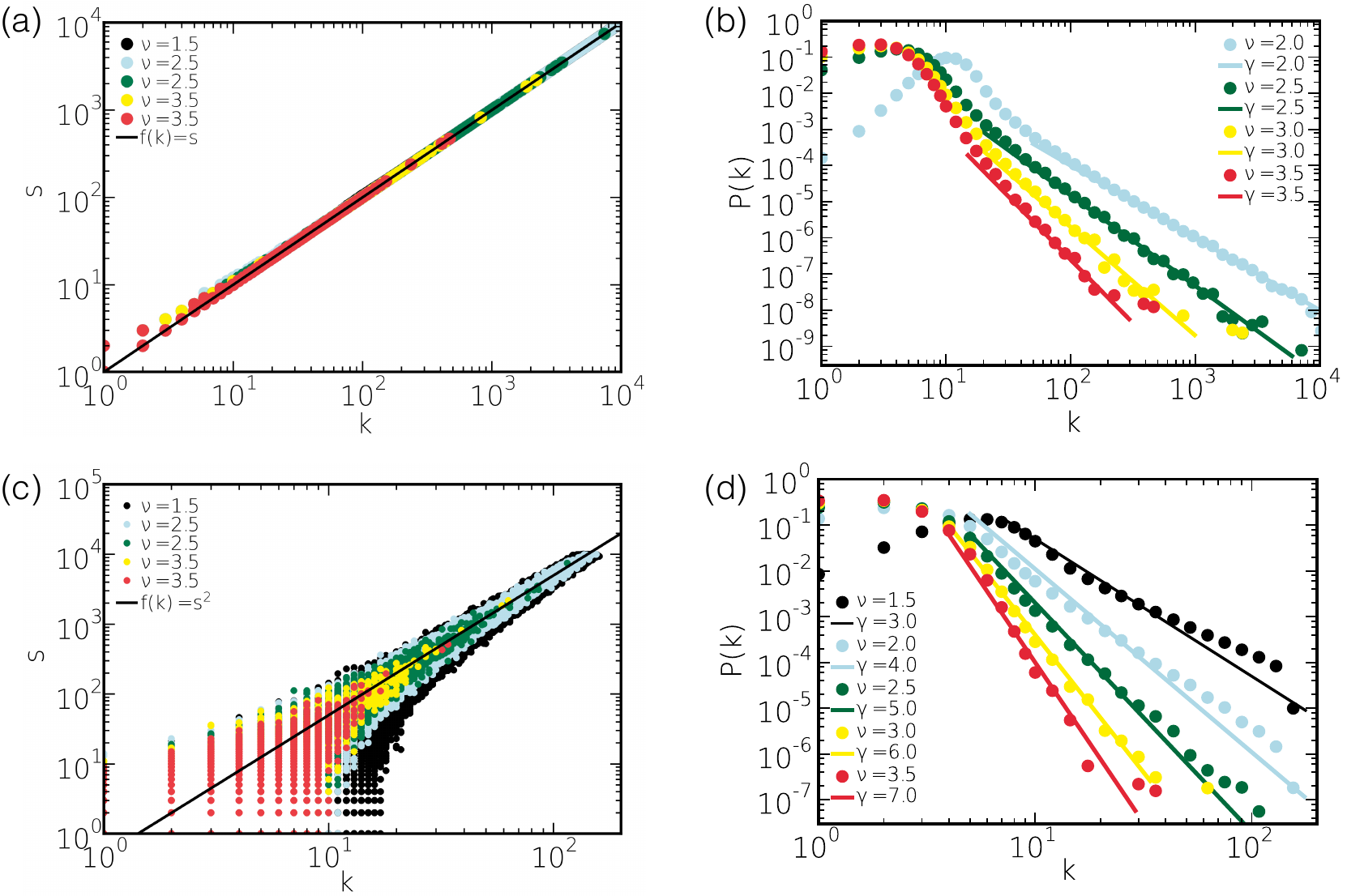}
\caption{Relation between node strength and degree in ML and RP model networks. (a) Degree-strength correlations for ML model in networks generated by different $\nu$ activity exponents. The fitted function is linear in $k$. (b) Degree distribution of the same ML networks. (c) Degree-strength correlations for RP model in networks generated by different $\nu$ activity exponents. The fitted function is $k^2$. (d) Degree distribution of the same RP networks.  Common parameters of the simulations: $N=1,000,000$, $m=1$, $\epsilon=0.001$, $T=10,000$.}
\label{fig:DAcorr}
\end{figure}

In the activity driven framework the degree of a node $i$ at time $t$ is decomposed into two parts as
\begin{equation}
k_i(t)=k_i^{out}(t)+k_i^{in}(t)
\end{equation}
where $k_i^{out}$ is the number of other nodes whom the node $i$ connected, while $k_i^{in}$ is the number of other nodes who connected node $i$ up to time $t$. Similar to degrees, the probability to have a degree $k$ of a node $i$ at time $t$ can be decomposed in two terms as:
\begin{equation}
P(t,k_i)=P_{out}(t,k_i)+P_{in}(t,k_i)
\end{equation}
where for an RP network (when $m=1$) the two probabilities can be written as
\begin{equation}
P_{out}(t,k_i)=a_i \left[ \frac{k_i-1}{k_i}P(t-1,k_i-1)+\frac{k_i-1}{k_i+1}P(t-1,k_i) \right]
\end{equation}
and
\begin{equation}
P_{in}(t,k_i)= \sum_{i\neq j}a_j\frac{1}{k_j+1}\frac{1}{(N-k_j+1)} P(t-1,k_j)P(t-1,e_{ij}\notin E_{t-1})
\end{equation}
where $E_t$ denotes the actual edge list at time $t$. These equations could provide us the relation between $\gamma$ and $\nu$, however no closed analytical solution has been found so far. 

Another way to estimate the relation between the activity and degree exponents in evolving RP networks is by directly measuring this correlation in large scale numerical simulations. In case of ML processes we have seen that $P(a)\sim P(k)$ \cite{Perra2012} thus a $a\sim k$ linear correspondence should be apparent. By definition the $a$ activity rate of a node is proportional to its $s$ strength, the total number of events the node participated during the process, thus for the ML case $s\sim k$ relations should be also satisfied. This correlation is confirmed in Fig.\ref{fig:DAcorr}.a for several $\nu$ exponent values, where the correlation between $k$ and $s$ is apparent to be linear 
\begin{equation}
\gamma = \nu.
\end{equation}

By following the same train of thought, if we measure the same correlation for RP networks it should also disclose the dependency between the actual activity and degree exponents. Calculations in Fig.\ref{fig:DAcorr}.c indicates a more dispersed distribution of correlation values between $s$ and $k$, however it suggest that their relation can be characterized as $s \sim k^2$  independently from the $\gamma$ exponent value. This dependency allows us to estimate a scaling relation between the activity and the emerging degree distributions as
\begin{equation}
P(a)=\int a^{-\nu} da \simeq \int k^{-2\nu} da = \int k^{-2\nu+1} dk
\end{equation}
which provides us the relation
\begin{equation}
\gamma\simeq2\nu-1.
\end{equation}

\subsection{Degree and reinforcement dependence}

\begin{figure}[ht!] \centering
  \includegraphics[width=140mm]{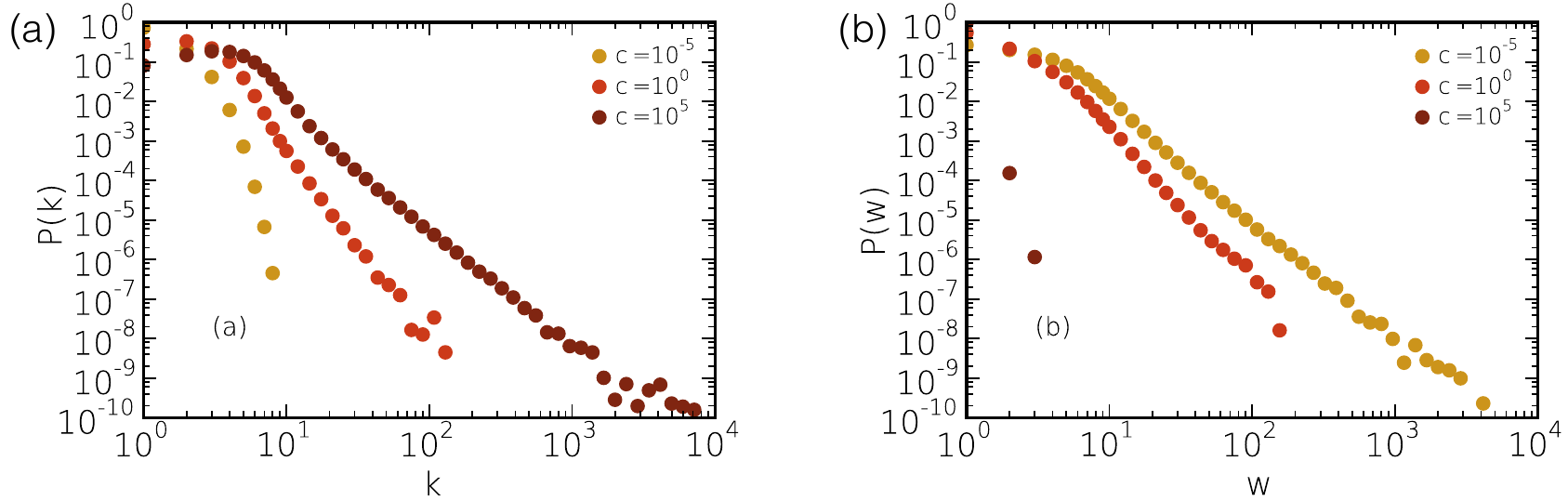}
\caption{The (a) degree distributions and (b) edge-weight distributions of RP networks with different reinforcement constant values $c=10^{-5}$, $c=10^{0}$ and $c=10^{5}$. Networks were generated with parameters $N=10^6$, $m=1$, $\gamma=2.8$, $\epsilon=0.0001$ and integrated through $T=10^4$ time steps.}
\label{fig:DCcorr}
\end{figure}

In reinforced activity-driven networks the decision of an agent to establish a new or reinforce an already existing connection is driven by a $p(n)$ probability. Earlier we showed that this probability can be well approximated with a simple analytical form as
\begin{equation}
p(n)=\frac{c}{n+c}
\end{equation}
where only the $c$ parameter depends on the activity and degree of the actual agent. Even in our model calculations (for simplicity) we fixed $c=1$ for every agent we remark that the evolving structural heterogeneities are depending on the choice of $c$. If $c\rightarrow 0$ the probability of calling a new friend goes to $0$, which reduce the emerging degree differences but increase the evolving weight heterogeneities. In the limiting case $P(k)$ becomes exponentially distributed while $P(w)\sim P(a)$ as it is shown in Fig.\ref{fig:DCcorr}.a and b for $c=10^{-5}$. On the other hand if $c \rightarrow \infty$ then $p(n)\rightarrow 1$ and the model approaches the memoryless activity-driven model process. In this limit $P(w)$ becomes and exponential distribution and $P(k)$ takes the same functional form as $P(a)$ as it is depicted in Fig.\ref{fig:DCcorr}.a and b for $c=10^{5}$. This way by varying $c$ one can control the emerging structural heterogeneities. One could even devise a model where $c$ follow a specific correlations with the activity of the actual agent, however, we let this kind of model extensions to be the subjects of future studies.

\section{Spreading rate dependencies}

During our simulations of rumour spreading processes we assumed that the actual values of $\lambda$ and $\alpha$  not, but their relative values matter for the equilibrium contagious level. To support this assumption we repeated some simulations of spreading processes with ordinary parameters and single random seeds. Here, instead of fixing $\lambda$ to unity, we performed measurements with $\lambda=0.8$, $0.6$, $0.4$ and $0.2$ values and record the $r_{RP}^{eq}(\alpha)/r_{ML}^{eq}(\alpha)$ curves in each case with $\alpha$ values ranging from $0$ to $\lambda$.

\begin{figure}[ht!] \centering
\includegraphics[width=80mm]{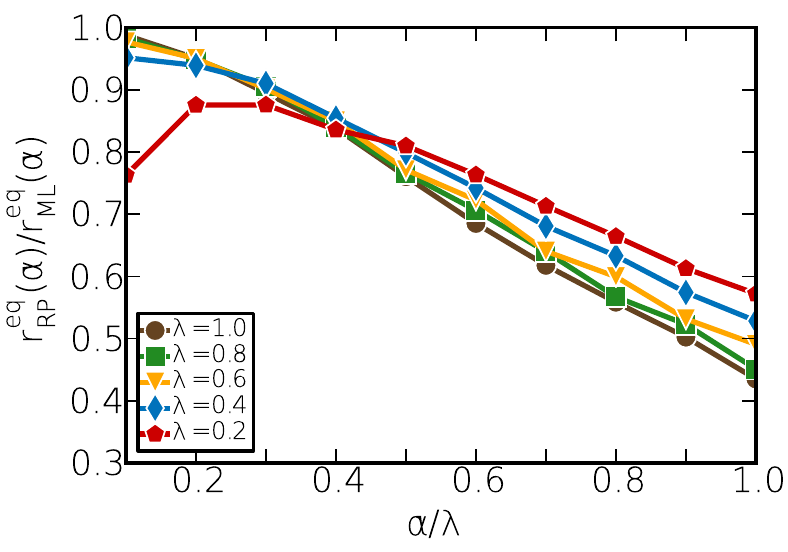}
\caption{The $r_{RP}^{eq}(\alpha)/r_{ML}^{eq}(\alpha)$ curves for rumour spreading processes with various $\lambda$ infection rate values. Simulations were performed for ML and RP processes with $\lambda=1.0$, $0.8$, $0.6$, $0.4$ and $0.2$ with relative $\alpha/\lambda$ values ranging between $0...1$. Results were averaged over $1000$ surviving simulations (see main text) with parameters $N=10,000$ , $m=1$, $\epsilon=0.001$, $T=50,000$.}
\label{fig:SIlambda}
\end{figure}

The results depicted in Fig.\ref{fig:SIlambda} demonstrate that only the relative values of the two rates matter for the equilibrium contagious level. Here curves corresponding to $\lambda=1.0$, $0.8$, $0.6$ and $0.4$ are very similar with differences only due random fluctuations (all of them were averaged over 1000 realizations). The largest discrepancy appears with $\lambda=0.2$  for small $\alpha$ values. This is because if $\lambda$ is small, the rumour spreads very slowly and even it would reach the same contagious level in equilibrium as any corresponding processes, it takes much longer time to reach this state. Since the time window was fixed to $T=50.000$ for every simulations, some processes could not arrive to the final equilibrium state which induced the discrepant scaling of $r_{RP}^{eq}(\alpha)/r_{ML}^{eq}(\alpha)$ ratios for smaller $\alpha/\lambda$ values.

\section{Surviving probability}

The surviving probability is defined as the probability that a system still contains agents in the spreader state at time $t$ (in other words the rumour survived up to $t$). This probability was measured in the MPC data-driven simulations and results were reported in the main text. Here we repeated these measurements for rumours spreading in activity-driven networks.

\begin{figure}[ht!] \centering
\includegraphics[width=120mm]{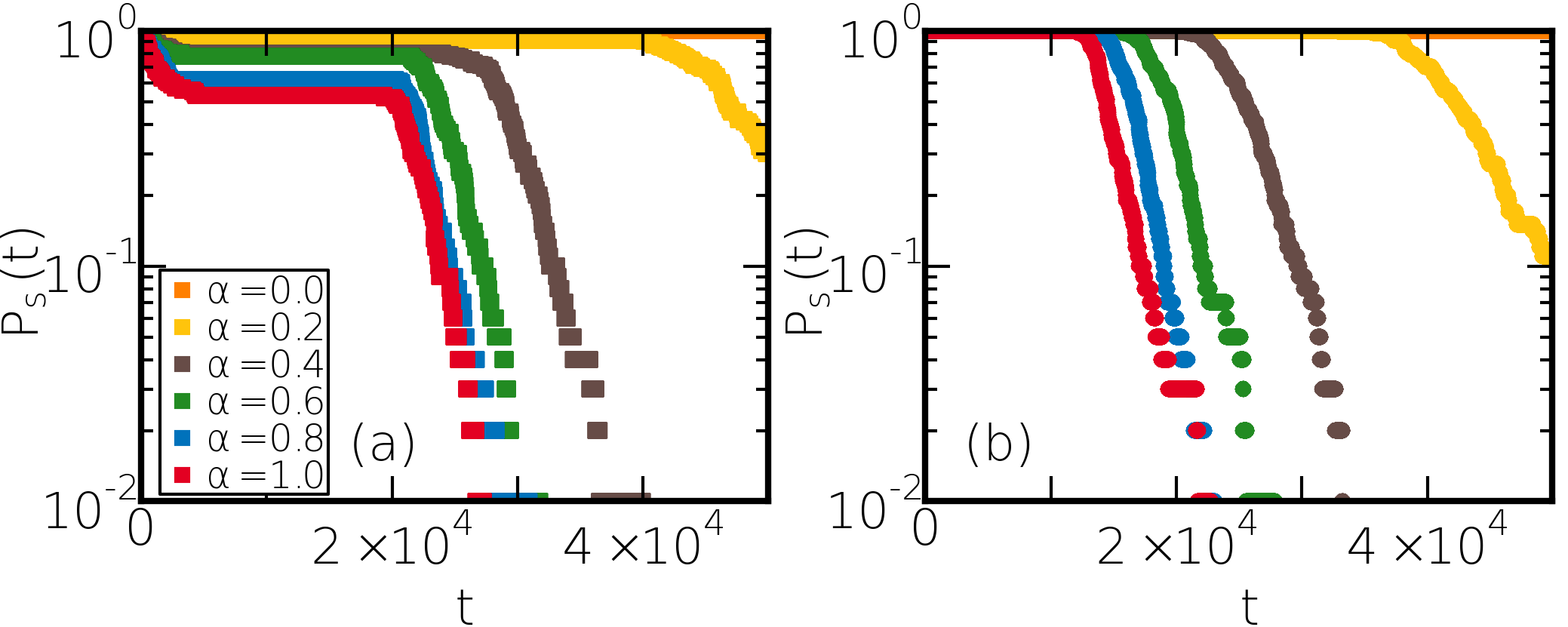}
\caption{Ps(t) Surviving probability of rumour spreading processes driven by (a) reinforced and (d) memoryless activity-driven processes. Probability values were calculated on networks with parameters $N=10^5$, $m=1$, $\epsilon=0.001$, $T=50,000$ for rumours initiated from a single random seeds. Results are averaged over $10^2$ realizations.}
\label{fig:SIPs}
\end{figure}

If the network evolution is driven by reinforced interaction processes (see Fig.\ref{fig:SIPs}.a) we recover the same effect what was observed during the data-driven simulations (see Fig.7.c in the main text). Repetitive interactions and memory play apparent roles in the early stage of the spreading processes as the rumour may die out shortly after its initiation and could spread only locally. This behaviour can be concluded from Fig.\ref{fig:SIPs}.a (and simultaneously from Fig.7.c in the main text) where the surviving probability rapidly decrease in the initial time regime for larger $\alpha$ values. After the rumour survives the initial stage, it spreads globally and reach a considerable fraction of the network.

On the other hand if the activity-driven process is memoryless (or equivalently if the data-driven spreading is evolving on shuffled event sequences) no repetitive interactions effect the initial temporal regime of the process and the rumour spreads always globally as it is evidenced Fig.\ref{fig:SIPs}.b (and in Fig.7.d in the main text where). This qualitative match between the model process and the data-driven simulations provides further prove that our model captures the role of memory and reinforcement processes in a consistent way.

\section{Effect of memory on spreading processes}

We discussed in the main text that memory processes play a strong influence on the emergence of any collective phenomena. Their effect can be two-fold. First of all, the repeated interactions slow down the evolution of the largest connected component of the emerging network. In addition they also play a role how influence can pass between nodes via temporal interactions. Here we demonstrate these effects by executing simulations of susceptible-infected (SI) processes on the evolving temporal ML and RP driven model networks and on their static  integrated structure. Note that the SI process by definition is one extreme case of the SIS, SIR and the rumour spreading models (discussed in the main text) if we choose the recovery rate $\alpha=0$.

\begin{figure}[ht!] \centering
\includegraphics[width=120mm]{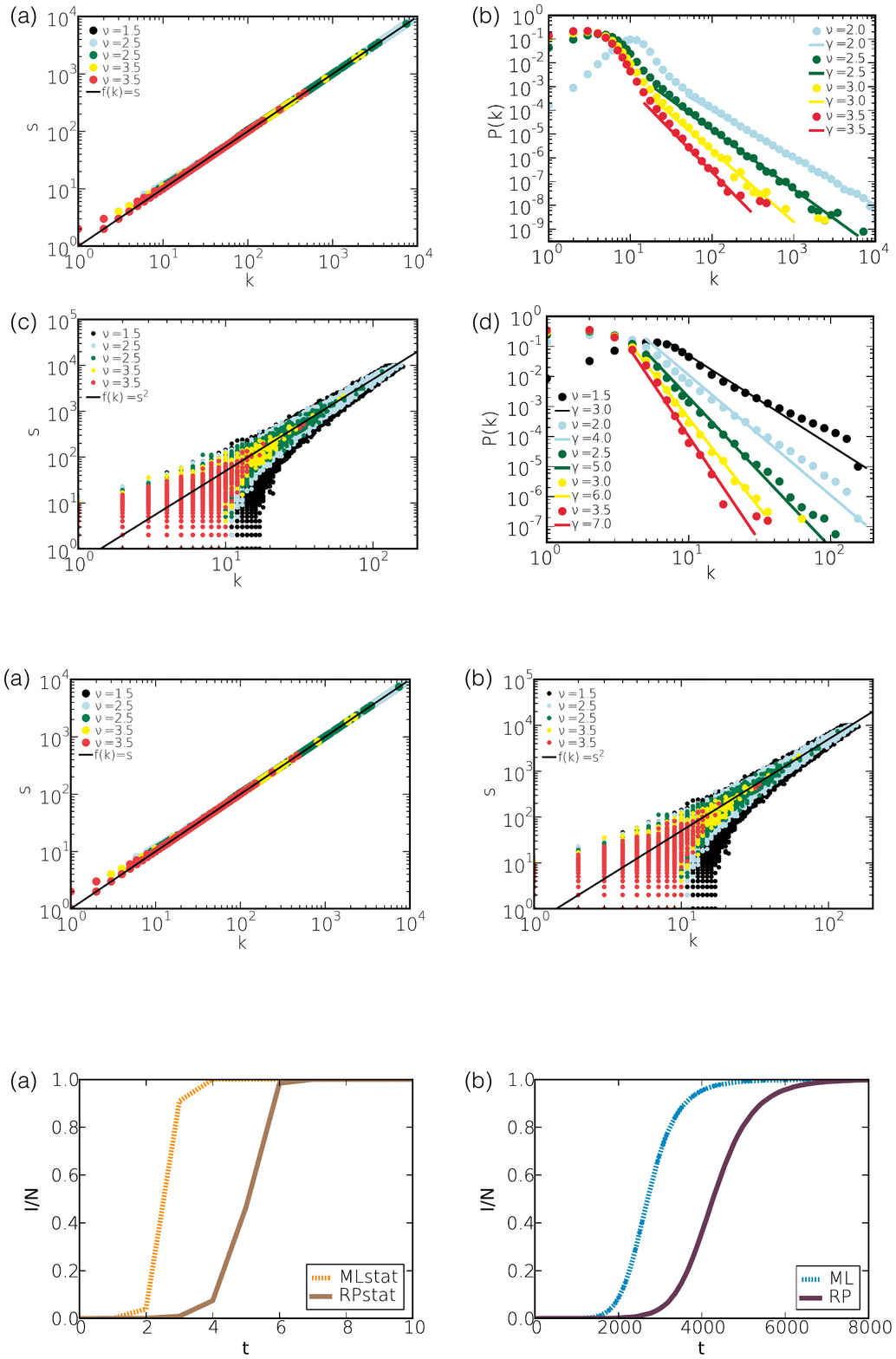}
\caption{Effects of memory on SI spreading processes. In panel (a) we show the $I/N$ fraction of infected nodes as a function of time evolving on static networks generated by ML (yellow dashed line) and RP (brown solid line) processes. Panel (b) depicts similar curves executed with the same parameters but where the infection process was co-evolving with the ML (blue dashed line) and RP (purple solid line) driven temporal model networks. Each simulations were executed with $N=10^5$ nodes for $T=10^4$ and with $\lambda=1$ infection rate (all other parameters are described in MM).}
\label{fig:SISIt}
\end{figure}

In case of static networks first we integrated the structure for $T$ time steps, then we induced the infection to a randomly selected seed node, and executed the measurement until the infection rate reached $I/N=1$. The spreading curves in Fig.\ref{fig:SISIt}.a demonstrate the structural effect of memory. In both ML and RP cases the infection reached every node in a few iteration steps, however due to the sparse connectivity of the static RP structure it evolved slightly slower in this case.

Strikingly larger different behaviour emerging in the temporal case. Here the infection was induced in the beginning of the process and it was allowed to co-evolve with the emerging structure. The infection could spread only via temporal connections, which slowed down the speed of infection with three orders of magnitudes compared to the static cases. In addition the repeated interactions reduced further the speed of spreading in case of the RP driven dynamics, which needed approximately two times more iterations to reach every node in the network.

\end{document}